\documentclass[a4paper,11pt]{article}

\usepackage{ifthen}
\usepackage{psfig}
\usepackage{url}
\usepackage{latexsym}
\usepackage{amssymb}
\usepackage{amsmath}
\usepackage{graphicx}
\usepackage{eepic}
\usepackage[all]{xy}
\usepackage{color}

\newcommand{\SetFigFont}[3]{\footnotesize} 

\usepackage{ifthen}

\ifthenelse{\isundefined{\definition}}{\newtheorem{definition}{Definition}}{}

\ifthenelse{\isundefined{\theorem}}{\newtheorem{theorem}{Theorem}}{}
\ifthenelse{\isundefined{\lemma}}{\newtheorem{lemma}{Lemma}}{}
\ifthenelse{\isundefined{\corollary}}{\newtheorem{corollary}{Corollary}}{}
\ifthenelse{\isundefined{\example}}{\newtheorem{example}{Example}}{}
\ifthenelse{\isundefined{\property}}{\newtheorem{property}{Property}}{}
\ifthenelse{\isundefined{\proof}}{\newenvironment{proof}{\noindent{\bf Proof.}\hspace{0.05in}}{\par\mbox{}\par}}{}

\newcommand{\vs}{\vspace{0.3cm}}

\ifthenelse{\isundefined{\qed}}{\newcommand{\qed}{\mbox{$\blacksquare$}}}{}

\newenvironment{lists}[1]{
                 \begin{list}{}{
                     \setlength{\listparindent}{0in}
                     \settowidth{\labelwidth}{#1}
                     \setlength{\leftmargin}{\labelwidth}
                     \addtolength{\leftmargin}{\labelsep}
                     }
                 }{
                 \end{list}
                 }

\newenvironment{given-find}[2]{
                               \vs 
                               \noindent \hrule
                               \begin{lists}{Given:XX}
                               \item[\sc Given: \hfill] #1                                 
                               \item[\sc Find: \hfill] #2                               
                               \vs 
                               \noindent \hrule 
                               }{
                               \end{lists}
                               }

\newcommand{\SDen}{\mathbb{S}}
\newcommand{\STrue}{\mathbf{T}}
\newcommand{\SFalse}{\mathbf{F}}

\newcommand{\SExt}[1]{\mbox{Ext}_{#1}}

\newcommand{\SIntL}[1]{\underline{#1}}
\newcommand{\SIntR}[1]{\overline{#1}}
\newcommand{\SInt}[1]{[\underline{#1}, \overline{#1}]}

\newcommand{\SPow}[1]{2^{#1}}

\newcommand{\SMG}{\SV^{*}}  
\newcommand{\SQC}{\SQA^{*}} 

\newcommand{\SymDomain}{{\cal D}}

\newcommand{\SStruc}{{\cal S}}
\newcommand{\SVarAss}{\theta}

\newcommand{\SKAp}[1]{\Hat{#1}}
\newcommand{\SBAp}[1]{\Tilde{#1}}
\newcommand{\SSubs}[3]{#1\frac{#2}{#3}}

\newcommand{\lsem}{[\![}
\newcommand{\rsem}{]\!]}
\newcommand{\SBIS}[2]{\lsem{#1},{#2}\rsem}   
\newcommand{\SBI}[1]{\SBIS{\underline{#1}}{\overline{#1}}}   
\newcommand{\SQA}{q}
\newcommand{\SV}{\mu}

\title{Convergent Approximate\\Solving of First-Order Constraints\\by Approximate Quantifiers}

\author{STEFAN RATSCHAN\\Research Institute for Symbolic Computation\\
Universit{\"a}t Linz}

\begin{document}

\maketitle

\begin{abstract}
  Exactly solving first-order constraints (i.e., first-order formulas over a certain
  predefined structure) can be a very hard, or even undecidable problem. In continuous
  structures like the real numbers it is promising to compute approximate solutions
  instead of exact ones.  However, the quantifiers of the first-order predicate language
  are an obstacle to allowing approximations to arbitrary small error bounds.  In this
  paper we remove this obstacle by modifying the first-order language and replacing the
  classical quantifiers with approximate quantifiers.  These also have two additional
  advantages: First, they are tunable, in the sense that they allow the user to decide on
  the trade-off between precision and efficiency.  Second, they introduce additional
  expressivity into the first-order language by allowing reasoning over the size of
  solution sets.
\end{abstract}

\section{Introduction}

Solving first-order constraints, (i.e., first-order formulas over a
certain predefined structure), and especially first-order constraints over the reals. 
has numerous
applications~\cite{Ratschan:01e,Dorato:95,Benhamou:00a,Dolzmann:97,Hong:97,Jirstrand:96}.
However, solving such constraints over the reals is either highly complex (e.g.,
when considering the predicate symbols $=$ and $\leq$, and the function symbols $+$ and
$\times$~\cite{Tarski:51,Fischer:74,Davenport:88,Weispfenning:88}), or
impossible~\cite{Tarski:51,Richardson:68}.
To deal with this problem, one can either restrict one-selves to more special problem
classes (see e.g.~\cite{Weispfenning:94,Hong:93a,GonzalezVega:96}), or relax the problem
by allowing approximation up to a user-specified error bound (as proposed by
H.~Hong~\cite{Hong:95c}). This paper studies the general feasibility of the second
approach. Its main contributions are: To show that even for this relaxed specification
we might have to do exact intermediate computation; and to introduce a modification of the
first-order predicate language---approximate quantifiers---for which this problem does
not occur. 


These quantifiers have two additional advantages: First, they are tunable, in the sense
that they allow the user to decide on the trade-off between precision and efficiency.
Second, they introduce additional expressivity into the first-order predicate language, by
allowing reasoning over the size of solution sets.

The first step to introduce approximate quantifiers is to allow quantifiers with a
positive real annotation $\SQA$, with the intuitive meaning that a formula
$\exists_{\SQA} x\;\phi$ is true iff the volume of the solution set of $\phi$ is
greater than $\SQA$. We will see that this does not yet allow the computation of
approximate solution sets up to arbitrarily small, user-specified, error bounds.

So we allow quantifiers to be annotated with a real interval $\SInt{\SQA}$, with the
intuitive meaning that the exact annotation can be any element of $\SInt{\SQA}$. This
allows an algorithm to choose the most suitable value in $\SInt{\SQA}$ and we do not
  care which one. This means that a sentence in the language does not have one distinct
truth-value but has a set of possible truth-values~(cmp.  with~\cite{Hehner:84}
or~\cite{Hoare:98}). We will prove that, from a good enough approximation of the solution
sets of the atomic sub-constraints, we can always compute at least one of these
truth-values, and thus one can always attain an arbitrarily small error bound when
computing approximate solution sets.



Following the usual approach (see e.g.~\cite{Walicki:97}), one would implement such a
logic using sets of truth-values (representing a many-valued logic~\cite{Urquhart:86,Rescher:69})
instead of single truth-values. We show that this approach is not suitable here and
present a new method that is completely orthogonal to the semantics usually given to
formulas when we do not know the value of certain predicate and function symbols and
thus assign validity to the formula using all possible predicate and function symbol
assignments (i.e., interpretations). Although arising from problems over real numbers, the
resulting first-order language is completely domain-independent.

The above situation that there are several possible values for an object, and we either do
not know or do not care which one should be taken, is commonly called \emph{don't know}
and \emph{don't care nondeterminism}, respectively. Here both forms occur at the same
time, which creates various difficulties through their interaction. Our approach gives
general insight into such a situation by showing how one can compute with such
nondeterministic objects, nevertheless. So our language can be easily extended to take
into account nondeterminism coming from other sources. For example uncertain coefficients
of occurring polynomials can be either modeled as don't care nondeterminism (the
\emph{united} approach) or as don't know nondeterminism (the \emph{robust}
approach)~\cite{Bouchon-Meunier:96,Gardenes:01,Shary:96}.

The structure of the paper is as follows: In Section~\ref{sec:appr-constr-solv}, we give
the specification of solving first-order constraints approximately up to some
user-specified error bound, and informally show that this is impossible for constraints
containing classical quantifiers.  In Section~\ref{sec:model-appr-comp}, we develop a
suitable formal model for approximate computation. In Section~\ref{sec:prop-appr-solut}
use this model to formalize the approximate solving of first-order constraints and its problems discussed in
Section~\ref{sec:appr-constr-solv}.
In Section~\ref{chp:quant}, we give a
first-order language where the classical quantifiers are replaced by approximate ones.  In
Section~\ref{sec:appr-comp-appr}, we apply the formal model for approximately solving
first-order constraints to approximate quantifiers. In Section~\ref{sec:repr-appr-solut},
we show how to deal with the resulting two forms of nondeterminism. In
Section~\ref{sec:biappr-proj-conv}, we prove that one can solve constraints that contain
approximate instead of classical quantifiers up to an arbitrary small error bound. In
Section~\ref{sec:related-work}, we discuss related work, and in
Section~\ref{sec:conclusion}, we give a final conclusion.

\section{Approximate Solving of First-Order Constraints}
\label{sec:appr-constr-solv}


Throughout the paper we use the term ``constraint'' as a shortcut for ``first-order
constraint'', that is, a first-order formula over a certain, predefined structure
$\SStruc$.  We fix a set $V$ of variables and define a \emph{variable assignment} as a
function from $V$ to $\SStruc$. For a variable assignment $\SVarAss$, an element
$a\in\SStruc$, and variable $v\in V$, $\SSubs{\SVarAss}{a}{v}$ is the variable assignment
that is the same as $\SVarAss$ except that it assigns $a$ to $v$.

We define the property that a constraint \emph{is true} for a certain variable assignment
(or \emph{is satisfied by it}) as usual.  A \emph{potential solution set} is a set of
variable assignments (we use the adjective ``potential'' for signifying the
independence from a specific constraint), and the \emph{solution set of a first-order constraint
  $\phi$} is the set of variable assignments for which $\phi$ is true

Sometimes we denote the solution set of a
closed first-order constraint (i.e., \emph{sentence}) by the Boolean constant $\STrue$
(which represents the set of all variable assignments), or $\SFalse$ (which represents the
empty set). In this case we also speak of the \emph{truth-value} (instead of solution set)
of a first-order constraint.

Recall that the notion of volume is modeled in mathematics by measure spaces (see any
textbook on measure theory, for example~\cite{Halmos:50}, for details).  For any measure
$\SV$ and set $A$, the inner measure $\SIntL{\SV}(A)$ is the supremum of the measures of
all measurable subsets of $A$ (or $-\infty$, if this supremum does not exist), and the
outer measure $\SIntR{\SV}(A)$ is the infimum of the measures of all measurable supersets
of $A$ (or $\infty$, if this infimum does not exist). For any set $A$,
$\SIntL{\SV}(A)\leq\SIntR{\SV}(A)$. If $A$ is measurable, then
$\SV(A)=\SIntL{\SV}(A)=\SIntR{\SV}(A)$---in this case we often use the term \emph{volume}
for measure. Furthermore, we call a function $\SV$ \emph{possible measure} iff for any set
$A$, $\SV(A)\in [\SIntL{\SV}(A),\SIntR{\SV}(A)]$.  We straightforwardly extend any measure
on $\SStruc^{|V|}$ to a measure on (potential) solution sets by measuring the tuples corresponding to
the variable assignments in the (potential) solution set.

We want to solve constraints: Given some constraint, we want to get a simple (e.g.,
quantifier-free) representation of its solution set.
However, over the real numbers, this problem is either highly
complex~\cite{Tarski:51,Fischer:74,Davenport:88,Weispfenning:88} or
undecidable~\cite{Tarski:51,Richardson:68}.  So we can only hope to tackle the general
problem, if we relax it.  As proposed by H.~Hong~\cite{Hong:95c}, we do this by
allowing approximation. This results in the problem specification of
Figure~\ref{fig:spec}.

\begin{figure}[htbp]
\begin{given-find}{
A constraint $\phi$, and\\
a positive real error bound $\varepsilon$}{
Sets $Y$ and $N$ of variable assignments, such that\\
    $\phi$ is true for all elements of $Y$,\\
    $\phi$ is false for all elements of $N$,\\
    the volume of the variable assignments not in $Y$ or $N$ is smaller than $\varepsilon$
}\end{given-find}
   
    \caption{Problem Specification}
    \label{fig:spec}
\end{figure}

In this paper we assume that we already have an algorithm that implements this
specification for atomic input constraints. For example, one can use for this a
branch-and-bound approach based on interval arithmetic~\cite{Jaulin:01b}. In theory, one
would need arbitrary precision here. In practice, however, fast 
machine-precision floating-point arithmetic usually suffices.

We would like to find an algorithm that fulfills the given specification based on such a
solver for atomic constraints.  However, this is impossible in general, because sometimes
exact solution sets of atomic sub-constraints are needed to compute such approximate
solutions. We show this here informally and formalize the arguments in the two following
sections.
%
%
Take an input constraint of the form $\exists x\;\phi$ without free variables, where
$\phi$ is an atomic constraint with an empty solution set. Determine for all $x$, except
for a set of arbitrarily small but positive volume, whether $x$ is in the solution set of
$\phi$. From this information we cannot deduce that $\exists x\;\phi$ is false, because
some of the remaining $x$ might be in the solution set of $\phi$, in which case $\exists
x\;\phi$ would be true.

One could suspect that the reason for this problem is, that a classical quantifier has to
take into account arbitrarily small solution sets of the quantified sub-constraint. So let
us introduce an additional quantifier $\exists_{\SQA}$ (the \emph{existential volume
  quantifier}) into the first-order predicate language, where $\SQA$ is a non-negative
real number. A closed constraint of the form $\exists_{\SQA} x\;\phi$ is true iff the
volume of the $x$, for which $\phi$ is true, is greater than $\SQA$. But even for this
quantifier, we have the same problem: Take an input constraint of the form
$\exists_{\SQA} x\;\phi$.  Assume that the solution set of $\phi$ has exactly the volume
$\SQA$.  Determine for all $x$, except for a set of arbitrarily small but positive volume,
whether $x$ is in the solution set of $\phi$.  Again, from this information we cannot
deduce whether $\exists_{\SQA} x\;\phi$ is true.  Note that one can easily find
examples of constraints \emph{with} free variables, that show the same behavior; in this
case we can deduce for no element of the free-variable space whether it is an element of
the solution set of such a constraint.

Speaking in the language of numerical analysis, the problem of finding approximations of
solution sets of quantified constraints is \emph{ill-posed} for certain inputs.  This
means that one can only solve it by approximation methods either after using more
information, or after relaxing the problem.  Following the latter approach, we introduce
different quantifiers, for which the problem does not occur.

\section{A Model for Approximate Computation}
\label{sec:model-appr-comp}

Before solving the problems described in the last section, we formalize them. For this we
develop a model for approximate computation in this section, and apply it to approximate
solving of first-order constraints in the next section. Readers who want to
see a solution to the problem immediately, without being interested in formal proofs, can
skip these sections and can directly jump to Section~\ref{chp:quant}.

Assume a set $A$. Instead of doing exact computation in $A$, we use a set $\SKAp{A}$ for
approximating computation in $A$. For this we add a notion of error, that is, a function
from $\SKAp{A}$ to $\mathbb{R}^{+}$.  For example, one can do approximate computation for
the real numbers using the set of rational intervals. Here the error of an interval is its
width. Take the expression $2x+1$. If we know that $x$ is in the interval $[2,3]$ then we
can deduce that the value of the expression is within $[5, 7]$---a result with error $2$.

Here we would like to be able to make the output error arbitrarily small by making the input
error small:

\begin{definition}
\label{def:convergent}
  A function $f: \SKAp{A}\rightarrow \SKAp{A}$ is \emph{convergent} iff 
    for all $\varepsilon\in\mathbb{R}^{+}$
      there is a $\delta\in\mathbb{R}^{+}$ such that
        for all $\SKAp{a}\in\SKAp{A}$ such that the error of $\SKAp{a}$ is less or equal $\delta$,
          the error of $f(\SKAp{a})$ is less or equal $\varepsilon$.
\end{definition}

Note that this definition corresponds to the definition of uniform continuity in analysis;
this notion and similar ones are used as necessary conditions for computability in
effective analysis~\cite{Bishop:67,Bishop:85,Pour-El:89,Weihrauch:00}.

We assume that every element $\SKAp{a}\in\SKAp{A}$ gives us the information that the
result of some computation is in a certain subset of $A$.  So we use subsets of $A$ for
modeling approximate computation from now on and simply identify $\SKAp{A}$ with
$\SPow{A}$. As in the example of rational intervals, very often one just uses certain
subsets of $A$ that allow a convenient representation.

Observe that it can also occur that we cannot define a certain function exactly, but only
approximately. This means that there are several possible functions, and we do not know
which one is the correct one to use. Again we represent this \emph{don't know}
nondeterminism by a set of functions, which we call \emph{approximate function}. For
example, the function computing the weight of a certain mass depends on the constant
describing gravitational acceleration. The exact value of this constant depends on the
distance from the center of the earth, and so we have not exact value for it---only an
interval covering all its possible values on the surface of the earth. Therefore
several such functions are possible and the whole function is approximate.

If we want to do approximate computation, then the best possible result we can obtain
without producing wrong results, is:

\begin{definition}
  For an approximate function $\SKAp{f}$ such that each element is a function on $A$, for
  an $\SKAp{a}\in\SKAp{A}$
  \[ \SExt{\SKAp{f}}(\SKAp{a}):= \{ f(a)\:|\: f\in\SKAp{f}, a\in\SKAp{a} \} \] 
We call $\SExt{\SKAp{f}}$ the \emph{extension of $\SKAp{f}$}.
\end{definition}

For example, interval arithmetic~\cite{Moore:66,Neumaier:90} defines approximate
computation on the real numbers in this way. However, instead of computing with exact
rational end-points, the results are usually rounded to the smallest super-interval whose
end-points are floating-point numbers.

Clearly the fact that the extension of a function is not convergent, implies that we cannot
use approximate computation to compute this function up to an arbitrarily small error.

\section{Approximate Computation and Classical Quantifiers}
\label{sec:prop-appr-solut}


Now we use the tools developed in the last section to show that classical quantifiers are
an obstacle to approximate computation. Let $\SDen$ be the set of potential solution sets. Then we
can do approximate computation using $\SKAp{\SDen}$, whose elements we call
\emph{potential approximate solution sets}, where the \emph{approximate solution set of a constraint
  $\phi$} is a potential approximate solution set that contains the solution set of $\phi$.





We can deduce the information we want to find for our specification
(Figure~\ref{fig:spec}) as follows:

\begin{definition}
  A variable assignment $\SVarAss$ is \emph{determined to true} by a potential
  approximate solution set $\SKAp{d}$ iff for all $d\in\SKAp{d}$, $\SVarAss\in d$. It is
  \emph{determined to false} by $\SKAp{d}$ iff for all $d\in\SKAp{d}$, $\SVarAss\not\in d$.  It
  is \emph{determined} by $\SKAp{d}$ iff it is either determined to true or determined to
  false by $\SKAp{d}$.
\end{definition}

Now we can easily define the error of a potential approximate solution set $\SKAp{d}$ as
the upper measure of the set of all variable assignments that are not determined by
$\SKAp{d}$.

For example, an approximate solution set of the constraint $x^2+y^2\leq 1$ might contain
all the potential solution sets that do not contain elements outside of the rectangle
$[-1,1]\times[-1, 1]$ (the solution set of $x^2+y^2\leq 1$ is one of these). In this case
all variable assignments assigning values outside of $[-1, 1]$ are determined to false,
and no variable assignment is determined to true. By measuring the size of the rectangle $[-1,
1]\times[-1, 1]$ we get the error $4$ of this approximate solution set.



For a constraint of the form $L(\phi_{1},\dots,\phi_{n})$, where $L$ is either a
quantifier and a variable (in this case $n=1$) or a connective, the solution set of the
total constraint is a function of the solution sets of the sub-constraints
$\phi_{1},\dots,\phi_{n}$. It is easy to show that the functions corresponding to
connectives are convergent in the sense of Definition~\ref{def:convergent}. So we will
concentrate on the case where $L$ contains an (existential) quantifier. In this case we have the
following function on potential solution sets (see~\cite{Ratschan:00} for the other
cases).

\begin{definition}
  For a variable $v\in V$, the \emph{$v$-projection operator} is a function $P$ on
  potential solution sets such that a variable assignment $\SVarAss\in P(d)$ iff there is
  an $a\in\SStruc$ such that $\SSubs{\SVarAss}{a}{v}\in d$.
\end{definition}



So, for a constraint $\exists x\;\phi$, we can compute an approximate solution set of the
total constraint from an approximate solution set of $\phi$, by applying $\SExt{\{ P \}}$,
where $P$ is the $x$-projection operator, to the approximate solution set of $\phi$.
Extending the above example to the constraint $\exists y\; x^2+y^2\leq 1$, we can use the
$y$-projection operator to compute an approximate solution set of this constraint. The
result contains all potential solution sets that do not assign values less than $-1$ or
greater than $1$ to $x$. This determines all variable assignments for which $x$ is not in
$[-1, 1]$ to false, and leaves all other variable assignments undetermined.

Using the argument from Section~\ref{sec:appr-constr-solv} it is easy to show that the
extension of the approximate function that contains the projection operator is not
convergent. Thus we cannot fulfill our specification (Figure~\ref{fig:spec}) for
constraints containing quantifiers. So we have to find a similar, but easier problem for
which we can fulfill the specification. We do this by introducing quantifiers that
approximate classical quantification, but result in a convergent projection operator.

\section{A First-Order Predicate Language with Approximate Quantifiers}
\label{chp:quant}


Classical quantifiers and volume quantifiers do not allow approximation up to an
arbitrarily small error bound because they discontinuously change from false to true. If
the size of the solution set of the quantified constraint is exactly at this point of
discontinuous change, then we cannot decide between true and false using approximate
computation.

\begin{figure}[htbp]
  \begin{center}
    \input{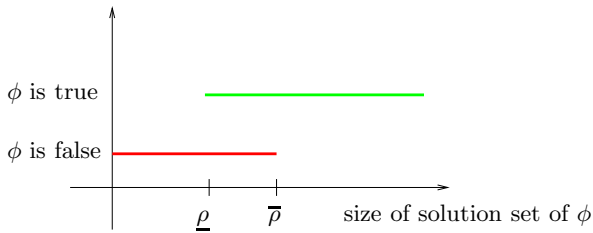}
    \caption{Approximate Quantifiers}
    \label{fig:apprQuant}
  \end{center}
\end{figure}

We avoid this discontinuous change by using volume quantifiers $\exists_{\SQA}$
($\forall_{\SQA}$) for which we do not specify the annotation $\SQA$ exactly, but only
require it to be within a certain interval $\SInt{\SQA}$.  Then a constraint solver can
choose an element of this interval element for which it can safely decide whether the
total constraint is true, using the available approximate information. This means that
sentences containing such an approximate quantifier can possibly be both true and
false---depending on which element of the interval $\SInt{\SQA}$ we choose. This happens
if the size of the solution set of a quantified constraint is within the interval $\SInt{\SQA}$ (see
Figure~\ref{fig:apprQuant}).

Recall that an approximate solution set of a constraint consists of several potential
solution sets, and we do not know which of them is the correct one---we have \emph{don't
  know} nondeterminism.  But here we are in exactly the dual situation: We allow several
equally possible truth-values, and we do not care, which of them is chosen---thus we
have \emph{don't care} nondeterminism.

Since our quantifiers depend on the size of the solution set of the quantified constraint,
we also have to deal with the situation when this solution set is not measurable.
Fortunately, this again is \emph{don't know} nondeterminism. If the solution set is not
measurable, then we only know that its volume is between the inner and the outer measure,
but we do not know which of these.

A naive approach to modeling such a situation, where a logical formula can have more than
one solution set, would propagate sets of truth-values by applying the logical symbols
element-wise.  For example, if for a sentence $\phi_{1}\wedge\phi_{2}$ the approximate
truth-value of $\phi_{1}$ is $\{\STrue\}$ and the approximate truth-value of $\phi_{2}$
is $\{\STrue,\SFalse\}$ then the combination of all these elements yields a truth-value
of $\{\STrue,\SFalse\}$ for the whole sentence.  This results in a many-valued
logic~\cite{Urquhart:86,Rescher:69}.  In order to show that this approach is not feasible,
we first demonstrate, that this would need a more complicated many-valued logic, and then
give a reason, why to avoid many-valued logics altogether:

One would need a more complicated many-valued logic, because of the need to consider the
interaction between the \emph{don't care} nondeterminism resulting from the approximate
quantifiers and the \emph{don't know} nondeterminism resulting from unmeasurable sets.
Approximate quantifiers can result in an approximate truth-value $\{ \STrue, \SFalse \}$.
An unmeasurable solution set can result in an unknown solution set of a formula.
Typically one would model this by an empty set.  But one wants to be able to assign the
approximate truth-value $\{\SFalse\}$ to the whole formula $\phi_{1}\wedge\phi_{2}$, if
$\phi_{1}$ has the approximate truth-value $\{\SFalse\}$ and $\phi_{2}$ has the empty
approximate solution set, although element-wise combination yields the empty set in this
case. This problem arises because a non-existing truth-value for $\phi_{2}$ means that it
can have two possible truth-values ($\STrue$ or $\SFalse$).  We do not know which one, but
we have modeled only the \emph{don't care} form of nondeterminism.

One could construct a many-valued logic that solves this
problem~\cite{Ginsberg:88,Fitting:91}, but there
is another problem that makes us avoid such an approach altogether: For example, consider
$\phi\wedge\neg\phi$, where $\phi$ has the approximate solution set $\{ \STrue, \SFalse
\}$. Then element-wise combination yields the approximate solution set $\{ \STrue,
\SFalse\}$, although we want this formula to be false in any case. The reason is, that a
many-valued logic forgets the information about the equality of $\phi$ in both branches of
$\wedge$~\cite{Urquhart:86,Rescher:69}. This also makes it impossible to define
$\leftrightarrow$ (equivalence) as an abbreviation. A similar problem also occurs in
interval mathematics~\cite{Moore:66,Neumaier:90}, where the information about equal terms
is lost.

Before going into the details of our solution, we fix the syntax of the new language.  It
is the usual one of the first-order predicate language, with the only exception that
instead of classical quantifiers, approximate quantifiers are used.  Consider the example
$\forall^{1}_{[0.1, 0.2]} x\;\exists^{2}_{[0.0, 0.1]} y\; [\;x>0 \wedge y=0\;]$.  The
quantifiers have a subscript consisting of a nonnegative real interval (the
\emph{annotation}).  Furthermore they have a positive integer superscript (the
\emph{tag}). We require that, within a formula,
quantifiers that have the same tag, also have the same annotation.

Tag equality indicates equal (nondeterministic) behavior of the according quantifiers.
That is, for quantifiers that have the same tag, the same element of the corresponding annotations
should always be chosen. This will allow us to make $A\wedge\neg A$ false in any case, and
makes the definition of $A\leftrightarrow B$ as an abbreviation for $(A\vee\neg B) \wedge
(\neg A\vee B)$ possible.

If the annotation of a quantifier is a one-element interval then we say that the
quantifier is \emph{deterministic}. A formula where all the tags of quantifiers that
are not deterministic, are different, is called \emph{free}. Often we do not explicitly
write down the tags but assume an arbitrary tagging such that a formula is free.

For assigning semantics to such formulas, we need to extend some of the usual definitions of the
first-order predicate logic:

\begin{definition}
  An \emph{m-structure} consists of a measure space $\SymDomain$, and for each relation and
  function symbol an according relation and function in $\SymDomain$. 
\end{definition}


As usual, by abuse of notation, we denote by $\SymDomain$ also the set on which the
measure space is defined.  
Terms can be
interpreted as usual in m-structures. We again fix an arbitrary m-structure $\SStruc$ with
measure space $\SymDomain$ that defines a measure $\SV$. For defining the semantics of approximate
quantification we use a method that is completely orthogonal to the semantics usually
given to formulas when we \emph{don't know} the value of certain predicate and function
symbols and thus assign validity to the formula using all possible predicate and function
symbol assignments.

It is straightforward to give semantics to a first-order constraint with approximate
quantifiers if we already know for each occurring quantifier with annotation
$\SInt{\SQA}$, which volume quantifier $\exists_{\SQA}$ or $\forall_{\SQA}$, where
$\SQA\in\SInt{\SQA}$, and which possible measure to use.  So, in analogy to the notion of
structure which assigns information to predicate and function symbols, we assign
information to quantifiers as follows:

\begin{definition}
  Given a first-order constraint $\phi$, a function $\SQC:
  \mathbb{N}\rightarrow\mathbb{R}$ is a \emph{quantifier choice for $\phi$} iff for every
  $t\in\mathbb{N}$ that occurs as a quantifier tag in $\phi$, $\SQC(t)\in [\SIntL{\SQA},
  \SIntR{\SQA}]$, where $[\SIntL{\SQA}, \SIntR{\SQA}]$ is the annotation of the
  quantifiers occurring in $\phi$ that are tagged by $t$.
\end{definition}

In a similar way, we can assign to each tag the possible measure that should be used
(recall from Section~\ref{sec:appr-constr-solv} that in the case of measurable sets a
possible measure just assigns its measure, otherwise any value between the inner and the
outer measure).

\begin{definition}
  A \emph{measure guess} is a function \[\SMG: \mathbb{N}\rightarrow (2^{\SymDomain}\rightarrow [0, \infty]),\] such that for all $t\in\mathbb{N}$,
  $\SMG(t)$ is a possible measure.
\end{definition}

It is an easy exercise to define the solution set of a constraint $\phi$, for a certain
quantifier choice $\SQC$ for $\phi$, and a certain measure-guess $\SMG$. Now we model that
we \emph{don't care} for the quantifier choices and \emph{don't know} about the right
measure guesses. For this we introduce two notions corresponding to the notion that a
constraint is true. The essence is, that now different quantifier choices can be used, and
each choice can result in a different overall result:

\begin{definition}
\label{def:truthVal}
A first-order constraint $\phi$ \emph{is true} for a variable assignment $\SVarAss$ iff
there is a quantifier-choice $\SQA^{*}$ for $\phi$, such that for all measure-guesses
$\SV^{*}$, $\phi$ is true for $\SVarAss$, $\SQA^{*}$, and $\SV^{*}$.  A
first-order constraint $\phi$ \emph{is false} for a variable assignment $\SVarAss$ iff
there is a quantifier-choice $\SQA^{*}$ for $\phi$, such that for all measure-guesses
$\SV^{*}$, $\phi$ is not true for $\SVarAss$, $\SQA^{*}$, and $\SV^{*}$.
\end{definition}

As an example assume a sentence $\exists_{\SInt{\SQA}} x\;\phi$, where $\phi$ has a
measurable solution set $d_{\phi}$. If the volume of $d_{\phi}$ is less or equal
$\SIntL{\SQA}$ the constraint is false; if the volume of $d_{\phi}$ is greater than
$\SIntR{\SQA}$, then $\phi$ is true; but if the volume of $d_{\phi}$ is within
$\SInt{\SQA}$, $\phi$ is both true and false. So we have exactly the behavior sketched in
Figure~\ref{fig:apprQuant}.





\section{Approximate Computation and Approximate Quantifiers}
\label{sec:appr-comp-appr}

The question remains, whether we can solve first-order constraints that contain
approximate instead of classical quantifiers, up to an arbitrarily small error bound. For
this we
apply our model for approximate computation of Section~\ref{sec:model-appr-comp} to
approximate quantification. The agenda will be, first to show how to propagate \emph{all}
available approximate information, and then to show, how to infer from this the
information needed for our specification in Figure~\ref{fig:spec}.

In Section~\ref{sec:model-appr-comp} we represented the uncertainty about a solution set
by the notion of approximate solution set of a constraint.  In
addition to this, here we also have the \emph{don't know} nondeterminism resulting from
measure guesses. So an \emph{approximate solution set of a constraint $\phi$ under a
  quantifier choice $\SQC$} is a potential approximate solution set that contains all solution sets
of $\phi$ under this quantifier choice $\SQC$ and any measure guess $\SMG$.
But in addition to this
nondeterminism, the semantics of our language, as given in Definition~\ref{def:truthVal},
takes into account the \emph{don't care} nondeterminism resulting from quantifier choices. This
means that several different approximate solution sets can be equally valid, depending on
the actual quantifier choice taken. For modeling this situation we introduce a second
level of approximation:

\begin{definition}
  A \emph{potential biapproximate solution set} is a set of potential approximate solution
  sets.  A \emph{biapproximate solution set of $\phi$} is a set of approximate solution
  sets of $\phi$.
\end{definition}

Note that here, when dealing with \emph{don't care} nondeterminism, we approximate the
case of full information, by taking a subset instead of superset of all possible objects.


In general, also for functions one can have these two forms of nondeterminism: On the one
hand, we can have several functions where we \emph{don't know} which is the right one. On
the other hand, we can have several functions where we \emph{don't care} which one is
chosen.  Analogously to solution sets, we model the combination of both by sets of
approximate functions, which we call \emph{biapproximate functions}.

For typesetting reasons, instead of $\SKAp{\SKAp{A}}$ we use the notation $\SBAp{A}$ to
denote biapproximate objects.  Again we can deduce the information that we want to find
for our specification (Figure~\ref{fig:spec}) as follows:

\begin{definition}
\label{def:1}
  A variable assignment $\SVarAss$ is \emph{determined to true} by a potential biapproximate
  solution set $\SBAp{d}$ iff there is a potential approximate solution set $\SKAp{d}\in\SBAp{d}$
  such that $\SVarAss$ is determined to true by $\SKAp{d}$. It is \emph{determined to
    false} by $\SBAp{d}$ iff there is a potential approximate solution set $\SKAp{d}\in\SBAp{d}$
  such $\SVarAss$ is determined to false by $d\in\SKAp{d}$.  It is \emph{determined} by
  $\SBAp{d}$ iff it is either determined to true or determined to false by $\SBAp{d}$.
\end{definition}

Now we again define the error of a potential biapproximate solution set $\SBAp{d}$ as the outer
measure of the set of all variable assignments not determined by $\SBAp{d}$.

As for classical quantification also approximate quantification results in a function on
potential solution sets:

\begin{definition}
  For a variable $v\in V$, a nonnegative real number $\SQA$, and a possible measure
  $\SV$, the \emph{$(v, \SQA,\SV)$-projection operator  $P_{\SQA,\SV}^{v}$} is a function
 on potential solution sets such that $\SVarAss\in
  P_{\SQA,\SV}^{v}(d)$ iff $\SV(\{ a\in\SStruc \,|\, \SSubs{\SVarAss}{a}{v}\in d\})>\SQA$.
\end{definition}

For a variable $v$ and nonnegative real number $\SQA$, we call the set of all $(v,
\SQA,\SV)$-projection operators \emph{approximate $(v,\SQA)$-projection operator} and
denote it by $P_{\SQA}^{v}$. This is an approximate function. For a fixed $v$ and interval
$\SInt{\SQA}$ we call the set of all $(v, \SQA)$-projection operators, where
$\SQA\in\SInt{\SQA}$, \emph{biapproximate $(v, \SInt{\SQA})$-projection operator}. This is
a biapproximate function.






Now we compute biapproximate solution sets by applying the extension of the functions that
correspond to the logical symbols. 

\begin{lemma}
  Let $\exists_{\SInt{\SQA}}^{t} v\;\phi$ be a constraint, $\SKAp{d}_{\phi}$ an
  approximate solution set of $\phi$ under a quantifier choice $\SQA^{*}$ for
  $\exists_{\SInt{\SQA}}^{t} v\;\phi$, and $P=P_{\SQA^{*}(t)}^{v}$. Then
    $\SExt{P}(\SKAp{d}_{\phi})$ is an approximate solution set of
  $\exists_{\SInt{\SQA}}^{t}\;\phi$ under $\SQA^{*}$.
\end{lemma}


\begin{theorem}
  Let $\exists_{\SInt{\SQA}}^{t} v\;\phi$ be a constraint such that the tag $t$ does not
  occur in $\phi$, let $\SBAp{d}_{\phi}$ be a biapproximate solution set of $\phi$, and
  let $P$ be the biapproximate $(v, \SInt{q})$-projection operator. Then
  $\SExt{P}(\SBAp{d}_{\phi})$ is a biapproximate solution set of
  $\exists_{\SInt{\SQA}}^{t} v\;\phi$.
\end{theorem}

\begin{proof}
  Let $\SKAp{d}$ be an arbitrary but fixed element of $\SExt{P}(\SBAp{d})$.  We have to
  prove that $\SKAp{d}$ is an approximate solution set of $\exists_{\SInt{\SQA}}^{t}
  v\;\phi$.  By definition of extension we know that there is a
  $\SKAp{d}_{\phi}\in\SBAp{d}_{\phi}$ and a $\SQA\in\SInt{\SQA}$ such that
  $\SKAp{d}=\SExt{P_{q}^{v}}(\SKAp{d}_{\phi})$. By definition of biapproximate
  solution set, $\SKAp{d}_{\phi}$ is an approximate solution set of $\phi$ under a
  quantifier choice $\SQA^{*}_{\phi}$ of $\phi$. Let $\SQA^{*}$ be such that it is equal
  to $\SQA^{*}_{\phi}$ on all tags occurring in $\phi$, and let it assign $\SQA$ to $t$,
  which is possible since $t$ does not occur in $\phi$. Then, by the previous lemma,
  $\SExt{P_{\SQA^{*}(t)}^{v}}(\SKAp{d}_{\phi})=\SKAp{d}$ is an approximate solution set of
  $\exists_{\SInt{\SQA}}^{t} v\;\phi$ under $\SQA^{*}$.  \qed
  

\end{proof}

  

The condition on the tags is always fulfilled for free constraints. If it does not hold,
the resulting biapproximate solution set might contain elements that are no approximate
solution sets of the constraint. The reason for this is, that the extension does not take
into account tag equality in a similar way as interval arithmetic does not take into
account equality of variables~\cite{Moore:66,Neumaier:90}.




\section{Representing Biapproximate Solution Sets}
\label{sec:repr-appr-solut}

Before studying, whether biapproximate projection is convergent, we first study
biapproximate solution sets in more details.  Observe that different potential biapproximate
solution sets can contain exactly the same information we need to find for fulfilling our
specification in Figure~\ref{fig:spec}.  For example, if we have the biapproximate truth
value $\{ \{ \STrue \} \}$ of a sentence, then we know that the sentence is true.  But, if
we have the biapproximate truth-value $\{ \{ \STrue \}, \{ \STrue, \SFalse \} \}$, then we
have exactly the same information. So there is some interaction between the two forms of
nondeterminism that potential biapproximate solution sets do not explicitely take into account, and
we can divide the potential biapproximate solution sets into equivalence classes such that each
equivalence class element contains the same information.


In this section we show, that in many cases it suffices to implement functions on
potential biapproximate solution sets (e.g., projection operators) just on the above equivalence
classes. Since the set of equivalence classes has a much lower cardinality than the full set of
potential biapproximate solution sets, we can then find a representation that is better suited for
computer implementation.  Furthermore we gain a valuable tool for the subsequent proof
that the biapproximate $\SInt{\SQA}$-projection operators are convergent, and find
interesting insight in the interaction between \emph{don't know} and \emph{don't care} nondeterminism.

Before studying the general case of potential biapproximate solution sets, we start with the easier
case of potential approximate solution sets. The information we want to extract are the elements
determined to true, and the elements determined to false by a potential approximate solution set
$\SKAp{d}$. These are exactly the elements $\bigcap\SKAp{d}$, and all elements not in
$\bigcup\SKAp{d}$. This gives us the equivalence relation $\SKAp{d}_{1}\sim\SKAp{d}_{2}$
iff $(\bigcap\SKAp{d}_{1},\bigcup\SKAp{d}_{1})=(\bigcap\SKAp{d}_{2},\bigcup\SKAp{d}_{2})$.

  
Now let us also define an order $\leq$ on functions on potential approximate solution sets such that
$f_{1}\leq f_{2}$ iff for all $d$, $f_{1}(d)\subseteq f_{2}(d)$. Also here we can form an
equivalence relation on approximate functions by defining $\SKAp{f}_{1}\sim \SKAp{f}_{2}$
iff $(\min\SKAp{f}_{1},\max\SKAp{f}_{1})=(\min\SKAp{f}_{2},\max\SKAp{f}_{2})$. We call a function
$f$ on potential solution sets \emph{monotonic} iff $d_{1}\subseteq d_{2}$ implies
$f(d_{1})\subseteq f(d_{2})$. In a similar way we define an approximate or biapproximate
function to be monotonic iff all its members are monotonic.

For computing with the equivalence classes instead of their members, we need to know
whether all members of equivalence classes behave equally for function application. For
this we pick a $\emph{canonical representative}$ from each equivalence class and prove
that all the other class members behave in the same way as this representative.  We can
order potential approximate solution sets by the subset relation, and so we can also use interval
notation on them: $[\SIntL{d},\SIntR{d}]:= \{ d \:|\: \SIntL{d}\subseteq
d\subseteq\SIntR{d} \}$. Observe that for any potential approximate solution set $\SKAp{d}$, the
interval $r(\SKAp{d}):=[\bigcap\SKAp{d},\bigcup\SKAp{d}]$ is equivalent to $\SKAp{d}$, and
for any approximate function $\SKAp{f}$ the interval
$r(\SKAp{f}):=[\min\SKAp{f},\max\SKAp{f}]$ is equivalent to $\SKAp{f}$. So we take
$r(\SKAp{d})$ and $r(\SKAp{f})$ as the canonical representatives. Now we have the
following congruence property:


\begin{lemma}
  For a potential approximate solution set $\SKAp{d}$ with canonical representative $\SInt{d}$ and a
  monotonic approximate function $\SKAp{f}$ with canonical representative $\SInt{f}$,
  $r(\SExt{\SKAp{f}}(\SKAp{d}))=[\SIntL{f}(\SIntL{d}), \SIntR{f}(\SIntR{d})]$.
\end{lemma}

\begin{proof}
By definition, $r(\SExt{\SKAp{f}}(\SKAp{d}))$ is equal to 
\[ [\bigcap\SExt{\SKAp{f}}(\SKAp{d}), \bigcup\SExt{\SKAp{f}}(\SKAp{d})]\]

Now, by inserting the definition of $\SExt{\SKAp{f}}$ and by monotonicity of $\SKAp{f}$
\[ \bigcap\SExt{\SKAp{f}}(\SKAp{d}) = \bigcap \{f(d) \:|\: f\in\SKAp{f}, d\in\SKAp{d} \} = (\min\SKAp{f})(\bigcap\SKAp{d}) = \SIntL{f}(\SIntL{d}) \]
and 
\[ \bigcup\SExt{\SKAp{f}}(\SKAp{d}) = \bigcup\{f(d) \:|\: f\in\SKAp{f}, d\in\SKAp{d} \} = (\max\SKAp{f})(\bigcup\SKAp{d}) = \SIntR{f}(\SIntR{d}). \]
\qed
\end{proof}



The $(v, \SQA,\SV)$-projection operators are monotonic. Furthermore the representation of
each approximate $(v, \SQA$)-projection operator is the interval $[P_{\SQA,\SIntL{\SV}}^{v},
P_{\SQA,\SIntR{\SV}}^{v}]$. So we can compute the approximate projection
of a potential approximate solution set by just computing with the interval bounds of the
corresponding representations.

Now we study the general case of potential biapproximate solution sets.  In a similar way as
potential approximate solution sets, different potential biapproximate solution sets can yield the same
information. Also here the information we want to extract are the elements determined to
true, and the elements determined to false by a potential biapproximate solution set $\SBAp{d}$ (see
Definition~\ref{def:1}). These are the elements $\bigcup \{ \bigcap\SKAp{d} \,|\,
\SKAp{d}\in\SBAp{d} \}$, and all elements not in $\bigcap \{ \bigcup\SKAp{d} \,|\,
\SKAp{d}\in\SBAp{d} \}$.  Again, this gives us an equivalence relation
$\SBAp{d}_{1}\sim\SBAp{d}_{2}$ iff $(\bigcup \{ \bigcap\SKAp{d} \,|\, \SKAp{d}\in\SBAp{d}_{1}
\},\bigcap \{ \bigcup\SKAp{d} \,|\, \SKAp{d}\in\SBAp{d}_{1} \})= (\bigcup \{ \bigcap\SKAp{d} \,|\,
\SKAp{d}\in\SBAp{d}_{2} \}, \bigcap \{ \bigcup\SKAp{d} \,|\, \SKAp{d}\in\SBAp{d}_{2} \})$. In
a similar way we get an equivalence relation on biapproximate functions by
$\SBAp{f}_{1}\sim\SBAp{f}_{2}$ iff 
$(\max\{ \min\SKAp{f} \,|\, \SKAp{f}\in\SBAp{f}_{1} \},\min\{ \max\SKAp{f} \,|\,
\SKAp{f}\in\SBAp{f}_{1}\}) = (\max\{ \min\SKAp{f} \,|\,\SKAp{f}\in\SBAp{f}_{2} \},\min\{ \max\SKAp{f} \,|\, \SKAp{f}\in\SBAp{f}_{2}\})$.


Again we pick canonical representatives from the equivalence classes. For any elements
$\SIntL{x}$ and $\SIntR{x}$ in a domain with a partial order $\leq$, let $\SBI{x}:= \{ [
y, y \cup\SIntR{x} ] \,|\, y\in[\SIntL{x}\cap\SIntR{x},\SIntL{x}] \}$. We call such an object
\emph{biinterval} (algebraically speaking the resulting objects form a
bilattice~\cite{Ginsberg:88,Fitting:91}). By forming biintervals we stay within an
equivalence class:

\begin{lemma}
  Let $A$ be a set, let $\SBAp{a}$ an element of the corresponding set $\SBAp{A}$, and let
  $(\SIntL{a}, \SIntR{a})= (\bigcup \{ \bigcap\SKAp{a} | \SKAp{a}\in\SBAp{a} \}, \bigcap \{ \bigcup\SKAp{a} | \SKAp{a}\in\SBAp{a} \})$; then
  $\SBI{a}\sim\SBAp{a}$.
\end{lemma}

\begin{proof}
\begin{multline*}
\bigcup \{ \bigcap\SKAp{a}|\SKAp{a}\in\SBI{a} \} = 
\bigcup \{ \bigcap [ a, a\cup\SIntR{a} ] |a\in[\SIntL{a}\cap\SIntR{a}, \SIntL{a} ] \} =\\
\bigcup \{  a |a\in[\SIntL{a}\cap\SIntR{a}, \SIntL{a} ] \} =
\SIntL{a} = \bigcup \{ \bigcap\SKAp{a} | \SKAp{a}\in\SBAp{a} \}
\end{multline*}

In a similar way $\bigcap \{ \bigcup\SKAp{a}|\SKAp{a}\in\SBI{a} \} = \bigcap \{ \bigcup\SKAp{a} | \SKAp{a}\in\SBAp{a} \}$. \qed
\end{proof}

So again, we denote the canonical representative $\SBIS{\bigcup \{ \bigcap\SKAp{d} |
  \SKAp{d}\in\SBAp{d} \}}{\bigcap \{ \bigcup\SKAp{d} | \SKAp{d}\in\SBAp{d} \}}$ of
$\SBAp{d}$ by $r(\SBAp{d})$, and the canonical representative $\SBIS{\max \{ \min\SKAp{f}
  | \SKAp{f}\in\SBAp{f}\}}{\min \{ \max\SKAp{f} | \SKAp{f}\in\SBAp{f}\} }$ of $\SBAp{f}$
by $r(\SBAp{f})$. Also here we can compute with the representatives because of the
following congruence property:

\begin{theorem}
\label{thm:1}
  For a potential biapproximate solution set $\SBAp{d}$ and a monotonic biapproximate function
  $\SBAp{f}$, 
  $r(\SExt{\SBAp{f}}(\SBAp{d}))=\SBIS{\SIntL{f}(\SIntL{d})}{\SIntR{f}(\SIntR{d})}$, where
  $\SBI{d}=r(\SBAp{d})$, and $\SBI{f}=r(\SBAp{f})$.
\end{theorem}

\begin{proof}
By definition, $r(\SExt{\SBAp{f}}(\SBAp{d}))$ is equal to 
\[\SBIS{
    \bigcup \{ \bigcap \SKAp{d} | \SKAp{d}\in\SExt{\SBAp{f}}(\SBAp{d}) \}
  }{
    \bigcap \{ \bigcup \SKAp{d} | \SKAp{d}\in\SExt{\SBAp{f}}(\SBAp{d}) \}}\]



We have:
\begin{align*}
\bigcup \{ \bigcap\SKAp{d} | \SKAp{d}\in\SExt{\SBAp{f}}(\SBAp{d}) \} = && \text{def. of Ext}\\
\bigcup \{ \bigcap\SExt{\SKAp{f}}(\SKAp{d}) | \SKAp{f}\in\SBAp{f},\SKAp{d}\in\SBAp{d} \} =&& \text{def. of Ext.}\\
\bigcup \{ \bigcap \{ f(d) | f\in\SKAp{f}, d\in\SKAp{d} \} | \SKAp{f}\in\SBAp{f},\SKAp{d}\in\SBAp{d} \} = && \text{monot.}\\
\bigcup \{ (\min\SKAp{f})(\bigcap\SKAp{d}) | \SKAp{f}\in\SBAp{f},\SKAp{d}\in\SBAp{d} \} = && \text{monot.}\\
(\max \{ \min\SKAp{f} | \SKAp{f}\in\SBAp{f}\})(\bigcup \{ \bigcap\SKAp{d} | \SKAp{d}\in\SBAp{d}\}) = && \text{def. of r}\\
\SIntL{f}(\SIntL{d})
\end{align*}

For the dual case an analogous argument holds. \qed
\end{proof}

For a biapproximate $\SInt{\SQA}$-projection operator $P$, $r(P)$ is the
biinterval $\SBIS{P_{\SIntL{\SQA},\SIntL{\SV}}}{P_{\SIntR{\SQA},\SIntR{\SV}}}$. As a
consequence of the above theorem, for every potential biapproximate solution set $\SBAp{d}$ we can
compute the projection of its equivalence class by just working on the bounds $\SIntL{d}$
and $\SIntR{d}$ of $r(\SBAp{d})$ to get
$\SBIS{P_{\SIntL{\SQA},\SIntL{\SV}}(\SIntL{d})}{P_{\SIntR{\SQA},\SIntR{\SV}}(\SIntR{d})}$.

\section{Biapproximate Projection is Convergent}
\label{sec:biappr-proj-conv}





Now we are ready to prove that for approximate quantifiers we can compute approximate
solution sets of constraints up to arbitrarily small error bounds. For this we call a
potential biapproximate solution set that contains only measurable solution sets \emph{measurable}.

\begin{definition}
  Let $\SBAp{d}$ be a potential biapproximate solution set. For any variable assignment $\SVarAss$
  the \emph{$v$-error of $\SBAp{d}$ over $\SVarAss$} is the outer measure of all
  $a\in\SStruc$ for which $\SSubs{\SVarAss}{a}{v}$ is not determined by $\SBAp{d}$.
\end{definition}

We first prove that for single points in the free variable space
we can attain an arbitrarily small error.


\begin{lemma}
\label{lem:4}
For every measurable potential biapproximate solution set $\SBAp{d}$, for the
biapproximate $(v, \SInt{q})$-projection operator P with $\SIntL{\SQA}<\SIntR{\SQA}$, for
almost all variable assignments $\SVarAss$ such that $\SVarAss$ is not determined by
$r(\SExt{P}(\SBAp{d}))$, the $v$-error of $\SBAp{d}$ over $\SVarAss$ is greater than
$\SIntR{\SQA}-\SIntL{\SQA}$.
\end{lemma}

\begin{proof}
  Observe that a variable assignment $\SVarAss$ is not determined by a potential biapproximate
  solution set with representation $\SBI{d}$ iff $\SVarAss\not\in\SIntL{d}$ and
  $\SVarAss\in\SIntR{d}$.  Let $r(d)=\SBI{d}$ and let $\SVarAss$ be such that it is not
  determined by $r(\SExt{P}(\SBI{d}))$, which is
  $\SBIS{P_{\SIntL{\SQA},\SIntL{\SV}}^{v}(\SIntL{d})}{P_{\SIntR{\SQA},\SIntR{\SV}}^{v}(\SIntL{d})}$
  by Theorem~\ref{thm:1}.  So, by definition of projection operator, $\SIntL{\SV}(\{ a \,|\,
  \SSubs{\SVarAss}{a}{v}\in\SIntL{d} \})\leq\SIntL{\SQA}$ and $\SIntR{\SV}(\{ a \,|\,
  \SSubs{\SVarAss}{a}{v}\in\SIntR{d} \})>\SIntR{\SQA}$. Since $\SIntL{d}$ and $\SIntR{d}$
  are measurable, by Fubini's theorem, the sets $\{ a \,|\,
  \SSubs{\SVarAss}{a}{v}\in\SIntL{d} \}$ and $\{ a \,|\,
  \SSubs{\SVarAss}{a}{v}\in\SIntR{d} \}$ are  measurable for almost all $\SVarAss$.
  The $\SVarAss$ that are not determined by $\SBI{d}$ are represented by the set $\{ a\:|\:
  \SSubs{\SVarAss}{a}{v}\not\in\SIntL{d}\:\wedge\:
  \SSubs{\SVarAss}{a}{v}\in\SIntR{d}\}$. Clearly the volume of this set 
  is greater than $\SIntR{\SQA}-\SIntL{\SQA}$.  \qed
\end{proof}

Since we have shown in the last section that for computing our specification
(Figure~\ref{fig:spec}), we can apply the canonical representative of projection operators
instead of the projection operators themselves, we now prove:

\begin{theorem}
  For $\SIntL{\SQA}<\SIntR{\SQA}$, and the biapproximate $(v, \SInt{\SQA})$-projection operator
  $P$, $r \circ \SExt{P}$ restricted to measurable potential biapproximate solution sets is convergent.
\end{theorem}

\begin{proof}
  We have to prove that
    for all $\varepsilon\in\mathbb{R}^{+}$,
      there is a $\delta\in\mathbb{R}^{+}$ such that
        for every measurable potential biapproximate solution set $\SBAp{d}$ such that the error of $\SBAp{d}$ is less than $\delta$,
          the error of $r(\SExt{P}(\SBAp{d}))$ is less than $\varepsilon$.
          
  Choose $(\SIntR{\SQA}-\SIntL{\SQA})\varepsilon$ for $\delta$, and let
  $\SBAp{d}$ be an arbitrary but fixed measurable potential biapproximate solution set. 
  We derive a contradiction from the assumption that the error of $\SBAp{d}$ 
  is less than $(\SIntR{\SQA}-\SIntL{\SQA})\varepsilon$, but the error of
  $r(\SExt{P}(\SBAp{d}))$ is greater or equal $\varepsilon$.

  By Lemma~\ref{lem:4}, for almost every $\SVarAss$ that is not determined by
  $r(\SExt{P}(\SBAp{d}))$ the $v$-error of $\SBAp{d}$ over $\SVarAss$ is greater 
  than $\SIntR{\SQA}-\SIntL{\SQA}$. The measure of these  $\SVarAss$ is greater or equal
   $\varepsilon$. So, by Fubini's theorem, the error of $\SBAp{d}$ is greater than
   $(\SIntR{\SQA}-\SIntL{\SQA})\varepsilon$, which is contradiction to our assumption. \qed


\end{proof}

Note that, if unmeasurable sets occur, then we still can ensure that the projection
operator is convergent, by making the interval $\SInt{\SQA}$ big enough. In general, we
can make computation faster by increasing the size of the interval $\SInt{\SQA}$, that is,
by decreasing the required precision.


In a similar way as the projection operators for quantifiers, also for the other logical
symbols ($\vee$, $\wedge$, $\neg$) there are corresponding functions on potential solution
sets~\cite{Ratschan:00}, and for free constraints we can use their extension to propagate
the according potential biapproximate solution sets.  The functions corresponding to
conjunction and disjunction are monotonic, and for the function $N$ corresponding to
negation, $d_{1}\subseteq d_{2}$ implies $N(d_{1}\supseteq d_{2})$, which entails a
similar property as Theorem~\ref{thm:1}. So we can use biinterval representation also
here.  

As an example take the constraint $x^2+y^2\leq 1$ and the biinterval
$\SBIS{\emptyset}{S_1}$, where the potential solution set $S_1$ is the set of all variable
assignments that assign elements of $[-1, 1]$ to both $x$ and $y$. This determines all
elements outside of the rectangle $[-1, 1]\times [-1, 1]$ to false and leaves the other
elements undetermined.  Furthermore, take the constraint $x\geq 0$ and the biinterval
$\SBIS{\emptyset}{S_2}$, where $S_2$ is the set of all variable assignments. This leaves all
variable assignments undetermined. The solution set of a conjunction is the intersection
of the solution sets of the according sub-constraints, and intersection is monotonic. So by
Theorem~\ref{thm:1} we can  take the intersection of the corresponding biinterval elements,
which is $\emptyset\cap\emptyset$ and $S_1\cap S_2$, and get the biinterval
$\SBIS{\emptyset}{S_1}$ as a biapproximate solution set for $x^2+y^2\leq 1\wedge x\geq 0$.

One can easily show that the functions corresponding to $\vee$,
$\wedge$, and $\neg$ are convergent.  So we can compute approximate solution sets of free
constraints from approximate solution sets of their atomic sub-constraints, as described
in Section~\ref{sec:prop-appr-solut}.  Furthermore---as long as no unmeasurable solution
sets occur and provided that for all approximate quantifiers the left bound $\SIntL{\SQA}$
of the annotation is strictly smaller than the right bound $\SIntR{\SQA}$---we can
attain arbitrarily small, user-defined, error.

\section{Related Work}
\label{sec:related-work}

Languages for modeling nondeterminism in various other forms have been introduced within
the frame of abstract data types---see for
example~\cite{Walicki:97,Walicki:95,Hesselink:88,Hussmann:93}.  There one uses
nondeterministic specifications to either model nondeterminism occurring in reality, or to
abstract away unnecessary details of the behavior of a real or desired system; these
details might be specified later. Also in analysis nondeterminism has been modeled in
order to deal with unknown/uncertain knowledge, and a large amount of classical
analysis has been extended to this case~\cite{Aubin:90,Aubin:91}.

In contrast to the above cases, in our work the deterministic (exact) specification is
already given, and we introduce nondeterminism only later, in order to be able to do
approximate computation for a relaxed specification.  Furthermore we deal with two forms
of nondeterminism at the same time whereas the above approaches are always confined to one
form of nondeterminism.

The idea to allow several equally valid outputs (i.e., \emph{don't care} nondeterminism) to make
certain problems computable is frequently used when doing exact numerical computation on the
reals via potentially infinite representations~\cite{Brouwer:20,Wiedmer:80,Weihrauch:00}. 



Modifications of the first-order language that allow reasoning about the size of sets have
been studied coming from logic~\cite{Keisler:85}, or knowledge
representation~\cite{Bacchus:90,Bacchus:90a}, and they are a main topic in the area of generalized
quantifiers~\cite{Mostowski:57,Vaeaenaenen:97}. However, these languages do not allow any
nondeterministic choice of the size specification and they circumvent the problem of how
to deal with unmeasurable sets by allowing only expressions whose solution sets are
measurable.

\section{Conclusion}
\label{sec:conclusion}



For constraints containing classical quantifiers, information about approximate solution
sets of the atomic sub-constraints of a first-order constraint does not suffice to compute
abitrarily precise approximate solution sets of the whole constraints. We have provided a
remedy for this problem by replacing the classical quantifiers in the first-order
predicate language by approximate quantifiers.

In addition to enabling approximation algorithms, this gives us both expressive power in
reasoning and the possibility of tunable algorithms (algorithms where the user can decide
about the tradeoff between speed and precision). We have implemented such an algorithm---a
detailed description and analysis of this implementation will be published elsewhere.

The question remains, how big we should choose the intervals of the quantifier annotation.
Too small intervals can hamper the efficiency of constraint solving, while too big
intervals can disturb the information one is looking for. 



Most of this research was done as a part of the author's Ph.D. work. The author thanks his
Ph.D. advisor, Hoon Hong, for all the guidance.

\small

\bibliographystyle{abbrv}

\bibliography{sratscha,sratscha_own}

\end{document}